# Dark Universe and distribution of Matter as Quantum Imprinting: the Quantum Origin of Universe.

Ignazio Licata[a, a1], Gerardo Iovane[b], Leonardo Chiatti[d], Elmo Benedetto[b,c], Fabrizio Tamburini[e]

[a] ISEM, Institute for Scientific Methodology, Via Ugo La Malfa n. 153 90146, Palermo, Italy

[b] Department of Computer Science, University of Salerno, Via Giovanni Paolo II, 132, 84084 Fisciano (Sa), Italy

[c] Department of Engineering, University of Sannio, Piazza Roma 21, 82100 Benevento, Italy

[d] AUSL VT Medical Physics Laboratory Via Enrico Fermi 15, 01100 Viterbo (Italy)

[e] ZKM Karlsruhe, Lorentzstraße 19, 76135, Karlsruhe, Germany

## Abstract

In this paper we analyze the Dark Matter problem and the distribution of matter through two different approaches, which are linked by the possibility that the solution of these astronomical puzzles should be sought in the quantum imprinting of the Universe. The first approach is based on a cosmological model formulated and developed in the last ten years by the first and third authors of this paper; the so-called "Archaic Universe". The second approach was formulated by Rosen in 1993 by considering the Friedman–Einstein equations as a simple one-dimensional dynamical system reducing the cosmological equations in terms of a Schrödinger equation. As an example, the quantum memory in cosmological dynamics could explain the apparently periodic structures of the Universe while Archaic Universe shows how the quantum phase concerns not only an ancient era of the Universe, but quantum facets permeating the entire Universe today.

## Introduction.

## Return to the Foundations: the peaceful cosmological coexistence

Many contemporary studies about the formation and distribution of matter in cosmology are dominated by phenomenological approaches based on changes in the standard model [1]. In this paper, we explore two approaches that go in the foundational direction. By this term we mean the more general conditions that can be imposed in cosmology before the Planck era and inflation, assuming that Quantum Mechanics (QM) and General Relativity (GR) are indeed the correct frameworks to identify these conditions. In spite of its generality, this line of attack to the problem is very selective and full of information. We know, for example, that a "peaceful cosmological coexistence" implies the removal of the singularity and the introduction of an imaginary time, which is a precursor of the ordinary time, so as to replace the old image of the "thermodynamic balloon" with the introduction of a wave function of the Universe calculated through the Feynman path integral. This, as it is well known, is the essence of Hartle-Hawking proposal [2], and the thus calculated wave function of the Universe can satisfy the Wheeler-DeWitt Equation [3]. One of the most radical and recent proposals to the question is the theory of the Archaic Universe, developed by two of the authors starting from de Sitter Projective Relativity; it can be considered an extension of the Hartle-Hawking condition [4]. This theory was actually formulated to solve a different problem, that of the conciliation of Projective General Relativity (PGR) with the cosmological principle. This conciliation is not trivial because the PGR represents the extension of de Sitter relativity to the general non-empty case. According to this theory every observer, in any epoch, is placed at the chronological distance $t_0 \approx \pm 10^{18}$ s from the two sheets (past and future) of a de Sitter horizon. The problem arises from how to define a cosmic time consistently in these circumstances. The Archaic Universe is a hypothetical predecessor (not in a chronological but ontological sense, hence the adjective "archaic") of space-time. This precursor consists of the four-dimensional surface of a half of the five-dimensional hyper-sphere of radius $r = ct_0$. The equator of this half hyper-sphere lies on the hyper-plan $x_0 = 0$ and the intersections with the various plans at constant $x_0$, $0 \leq x_0 \leq ct_0$, constitute the various three-dimensional "parallels". It is imagined that the surface of the hemisphere is the seat of virtual processes originated on the equator, each of which ends on the parallel corresponding to a specific value of $x_0$. The processes ending on the parallel corresponding to a given value of $x_0$ are interpreted as virtual fluctuations of "duration" $x_0/c$ (and therefore, according to the indetermination principle, of energy $\hbar c/x_0$) of a pre-vacuum, precursor of the current vacuum. The formal temperature $\hbar c/kx_0$ is associated with these processes, where $k$ is the Boltzmann constant. The formation of the Universe is described as a set of "nucleation" processes starting from a "universal reservoir" consisting of the virtual processes "written" on the four-dimensional surface [5-8]. The characteristics of this nucleation, as we shall see, suggest interesting hypotheses about the nature of the dark matter (DM).

Another foundational approach is due to Nathan Rosen, the historical collaborator of Einstein, and is based on the so-called cosmological Schrödinger equation [9]. At first instance, talking about Schrödinger equation in cosmology it would seem absurd but it is not. For example, in a recent paper [10], the author shows that secular perturbations upon self-gravitating disks exhibit a mathematical similarity to a quantum scattering theory applying a time-dependent Schrödinger equation. It is well known that gravity induces quantum effects such as the evaporation of black holes and the absence of an accepted theory of quantum gravity remains a deep problem in theoretical physics. The canonical quantization of GR leads to the Wheeler-DeWitt equation introducing the so-called Superspace, an infinite-dimensional space of all possible 3-metrics. Rosen, instead, preferred to start his work from the classical cosmological equations using a simplified quantization scheme. Consider a homogeneous isotropic Universe. It is described by the Robertson-Walker line element and by assuming a standard perfect fluid matter; we deduce the following Friedman equations,

$$\begin{cases} \dfrac{\ddot{a}}{a} = -\dfrac{4\pi G}{3c^2}(\epsilon + 3p) + \dfrac{\Lambda c^2}{3} \\ \left(\dfrac{\dot{a}}{a}\right)^2 + \dfrac{kc^2}{a^2} = \dfrac{8\pi G}{3c^2}\epsilon + \dfrac{\Lambda c^2}{3} \quad (1) \\ \dot{\epsilon} + 3\left(\dfrac{\dot{a}}{a}\right)(\epsilon + p) = 0 \end{cases}$$

where the parameter $\dot{a}$ denotes a derivative with respect to the cosmic time $t$ of the scale factor $a$. For a complete description we also have to include an equation of state to characterize the thermodynamical conditions of matter, namely,

$$p = (\gamma - 1)\epsilon = (\gamma - 1)c^2\rho \quad (2)$$

where $p$ is the pressure, $\epsilon$ the energy density, with the parameter $\gamma$ found in the interval $1 \leq \gamma \leq 2$. In particular, $\gamma = 1$ characterizes the matter-dominated era and the value $\gamma = 4/3$ is found in the radiation-dominated era. The solution of the third equation of (1) is

$$\epsilon = \epsilon_0 \left(\frac{a_0}{a}\right)^{3\gamma} \quad (3)$$

where the subscript 0 means that the physical quantities are calculated at today's time. In [9] the author suggests to multiply the second relation in (1) by a factor $\frac{1}{2}ma^2$, where $m$ is the mass of the Universe, getting the following equation,

$$\frac{1}{2}m\dot{a}^2 - \frac{m}{6}\left(\frac{8\pi G}{c^2}\epsilon + \Lambda c^2\right)a^2 = -\frac{1}{2}mkc^2 \quad (4)$$

Moreover, the author interprets the cosmological equation as an energy conservation law, writing

$$T + V = E \quad (5)$$

and therefore the kinetic energy becomes

$$T = \frac{1}{2}m\dot{a}^2 \quad (6)$$

and the potential energy

$$V(a) = -\frac{m}{6}\left(\frac{8\pi G}{c^2}\epsilon + \Lambda c^2\right)a^2 \quad (7)$$

Finally, for the total energy we obtain

$$E = -\frac{1}{2}mkc^2 \quad (8)$$

By considering the Newtonian equation of motion

$$m\ddot{a} = -\frac{dV}{da} \quad (9)$$

and by defining the momentum of mass as

$$P = m\dot{a} \quad (10)$$

the Hamiltonian is

$$H = \frac{P^2}{2m} + V(a) \quad (11)$$

At this point, Rosen uses the standard first quantization procedure

$$P \to -i\hbar\frac{\partial}{\partial a}, \quad H \to i\hbar\frac{\partial}{\partial t} \quad (12)$$

obtaining the Schrödinger equation

$$i\hbar \frac{\partial \psi}{\partial t} = H\psi \quad (13)$$

where $\psi = \psi(a, t)$ is the wave function of the Universe.

These are two different lines of attack to the problem, but both can be considered a way to constrain the solutions of Wheeler-DeWitt equation and overcome the well-known interpretative problems related to a "Universe wave function". In the case of the Archaic Universe we select the boundary conditions that lead to the observed Universe avoiding the proliferation of cosmological models; Rosen equation, on the other hand, is like introducing a Feynman path built on the Robertson-Walker scenario. It is no coincidence that these two theories imply interesting consequences on the distribution of matter and on DM because both impose an "imprinting" ab initio on the dynamics in the time of the Universe. This review is structured as follows: sections 1-5 briefly recall the conceptual and mathematical premises of the Archaic Universe, the origin of inertia and the problem of DM. In sections 6 and 7 the Rosen theory and its consequences for the distribution of matter are developed.

## 1. The Archaic Universe, a brief description.

The term "pre-vacuum" used in the Introduction means the existence, on the hemi-spherical surface, of a special class of four-dimensional frames of reference with a "time" axis $x_0$, connected by three-dimensional roto-translations; the latter are isomorphic to transformations of the de Sitter group. In these special frames the virtual particles of the pre-vacuum have null momentum and therefore are considered "at rest". The observers placed at the origins of these frames do not experience any "wind of aether" associated with any virtual process. The basic idea in this framework is that the collapse of the quantum-mechanical wave-function is an objective process, consisting in the temporal localization of elementary particles on intervals in the order of $\theta_0 \approx t_0/N_D$, where $N_D \approx 10^{41}$ is the Dirac number [7]. For values of $x_0$ lower than $c\theta_0 \approx 10^{-13}$ cm, the collapses are inhibited, and therefore no real interaction occurs. For higher values, wavefunction collapses can occur; in other words, the pre-vacuum is unstable with respect to the temporal localization of its particles. The virtual fluctuations of "temporal" extension $x_0 > c\theta_0$ are then converted, from the collapse of their wave functions, into real particles. In other words, a process of nucleation of real matter occurs that "empties" the pre-vacuum, converting it into matter and ordinary vacuum. This nucleation, seen from the spatio-temporal domain, is the big bang. Space-time is not connected with the Archaic Universe dynamically, but through three distinct transformations of coordinates: first of all, a Wick rotation, which transforms the half hyper-sphere into a portion of hyperboloid; then a gnomonic projection of this hyperboloid on the tangent plane; finally, a scale contraction by a factor $\approx N_D$. These three operations transform the privileged frames of reference of the unique archaic space into a triple continuous infinity of spaces (one for each tangent point), connected by transformations of the de Sitter group (in a sandwich between two inverse scale transformations). These are the private space-times of the fundamental observers, i.e. the *global* frames of reference used by these observers to coordinate the events. It is to these different (but equivalent) "points of view on the world" that the gravitational equations of the PGR are applied. The conditions of homogeneity and "spatial" isotropy of the pre-vacuum induce the cosmological principle, which, thus, becomes a condition to be applied to the solution of these equations. Similarly to the conventional Friedmann cosmology, the distance scale $R(t)$ must be determined; $t$ is the cosmic time. The essential difference with respect to the conventional approach is, however, that the cosmological model is univocally determined and it is a flat model with a positive cosmological constant. This scheme does not require inflation, although it can be added.

The situation at the big bang can therefore be described as follows. Consider a fundamental observer O that emits (receives) a signal consisting of a particle P. In the private space-time of O (of radius ≈ $ct_0$ to the big bang) the propagation of the signal from P to O, or vice versa, is described by a translation on space-time characterized by a set of known parameters. The private space-time of P is instead connected to that of O by a de Sitter translation of parameters ≈ $N_D$ times larger (in a sandwich between two inverse scale transformations of factors respectively ≈ $N_D$, $1/N_D$). The gravitational equations define the metric of the private space-time of O as a function of the distribution of matter-energy in such space-time but the solutions of these equations, related to different values of O, are connected by a global transformation of coordinates on the de Sitter "public" space-time with radius $r = ct_0$. It is important to understand that the Archaic Universe does not "precede" the current Universe chronologically. Rather, it is a kind of boundary condition: the PGR equations are applied to the private space-time of a generic fundamental observer, and the private space-times of the various observers are connected by coordinate transformations of the Archaic Universe. In a sense, therefore, the archaic layer of physical reality is always present, though it cannot be directly probed. The situation is very similar to that of observers on the earth's surface put in the impossibility to travel. They can deduce the spherical shape of the Earth from various observations (curvature of the sea surface, shape of the Earth's shadow in the eclipses of the Moon, etc.), but they cannot verify it directly. The spherical shape of the Earth is in fact a global property that can be directly experimented only by traveling the Earth's surface or comparing the data coming from differently positioned observers. In the following sections we will see how, in a slight extension of this scenario, the dark matter can find its place as a "fossil" of the archaic pre-vacuum. We will limit our analysis to the exposure of the essential ideas on this topic, by referring these ideas to other works present in the literature [11-13] to obtain a comparison with the observational data. As we will see, the idea is that a specific quantization of the inertia formulated in the Archaic Universe leads, in private space, to the dependence of some local effects (the dark matter thickening) derived from the *global* constant $t_0$.

## 2."Archaic" origin of the inertia principle (and its violation)

In the previous presentation of the Archaic Universe, the parallel of coordinate $x_0$ (counted from the equator) of the hyper-sphere was considered isothermal, at the temperature $\hbar c/kx_0$ ($k$ is the Boltzmann constant). However, we can assume the possibility that on this parallel there is an excess of temperature $u$ (assumed positive-valued, as discussed below in the text), and a temperature gradient $\nabla u$. As we will see, the ratio between the maximum value of $u$ and the background value $\hbar c/kx_0$ (where $0 \leq x_0 \leq 10^{-13}3$ cm) is $\leq 10^{-39}$. Therefore, the generalization we are considering leaves the theoretical framework of the Archaic Universe, defined in the previous section, unchanged. In the origin of each of the privileged frames of reference taken into consideration in the previous section, the pre-vacuum is at rest for each value of $x_0/c$. But there may also be the more general case in which the motion of the particles of the pre-vacuum with respect to that origin (and in correspondence thereof) is at zero mean over long enough intervals of $x_0/c$. We assume that each of these frames is associated with a spherical spatial bubble of radius $\rho$ centered on its origin, concentric to a second bubble of radius $\delta \leq \rho$ within which the excess of temperature $u$ with respect to $\hbar c/kx_0$ is constant. We assume that $u$ is null for distances from the origin $x > \rho$, while it takes intermediate values for $\delta$

$\leq x \leq \rho$. These intermediate values will be defined by the static limit of the thermal conduction equation, i.e. by the Laplace equation $\Delta u = 0$. Let us now go to a point in the archaic space whose distance from the origin of the frame of reference is $x$, with $0 \leq x \leq \rho$. In this point it is possible to define the thermal energy $ku$, and the pulsation $\omega = 2\pi/T$ given by the relation $\hbar\omega = ku$. The fundamental assumption is that a material point in $x$ accelerates towards the origin, with an acceleration $\boldsymbol{a}$ given by the relation ($a = |\boldsymbol{a}|$):

$$\frac{da}{dx} = \pm \frac{4\pi^2}{T^2}, \quad T = T(x) \qquad (14)$$

where the sign of the second side of the equation is positive in the region $0 \leq x \leq \delta$, negative in the region $\delta \leq x \leq \rho$ (that is, the acceleration is maximum for $x = \delta$). It should be noted that this is equivalent, in relativistic terms, to assume an appropriate Newtonian static metric within the bubble of radius $\rho$ centered in $x = 0$. We will call this bubble a pre-vacuum "molecule". By defining the time period $T = h/ku$ we have $T = T_1$ for $0 \leq x \leq \delta$, $T = \infty$ for $x \geq \rho$. The values of $T$ in the spherical corona $\delta \leq x \leq \rho$ will be obtained from the solution of the Laplace equation. We note that positive values of the period $T$ (the only ones with physical meaning) imply $u > 0$, as requested. Before going on, let us consider $T_0$, a sort of "internal period" of the origin $x = 0$ of the frame of reference; for the connection of $T_0$ with $T_1$ we will return later on. Here, we hypothesize that in the Archaic Universe the origin of a frame of reference is endowed with a finite extension in $x_0/c$ equal to $\theta_0$, a typical scale of the temporal localization of elementary particles [8-14]. The random choice of a particular origin on the $x_0/c$ axis thus corresponds to an a priori probability equal to the ratio $\theta_0/t_0$ between the interval $\theta_0$ and the extension of the $x_0/c$ axis, and an information $1/\alpha = -\log_2(\theta_0/t_0)$. If the entropy

$$\left[\frac{\theta_0}{t_0}\right] \log_2\left(\frac{t_0}{\theta_0}\right) = \frac{\theta_0}{\alpha t_0} \qquad (15)$$

is interpreted as an action (in units of the Planck's constant $h$) expressed in the interval $\theta_0$, it is possible to define an energy as a ratio of this action and $\theta_0$; this ratio is, in conventional units, $h/\alpha t_0$. It is then possible to define a new action as a product of $h/\alpha t_0$ for an interval of extension $T_0$ of the variable $x_0/c$. If we assume that this action is an integral multiple of $h$, we have $T_0 = \nu\alpha t_0$, with $\nu = 0, 1, 2,\ldots$ . In other works we have identified $\theta_0$ with the time necessary for light to travel the classical radius of the electron, and $\alpha$ with the fine structure constant [14]. It should be noted that in the "Einsteinian" limit of an infinite de Sitter time, that is for $t_0 \to \infty$, a finite (and undetermined) value of $T_0$ is only possible for $\nu = 0$. The case $\nu = 0$ corresponds to the unexcited pre-vacuum; its particles are at rest with respect to the origins of a continuum of frames of reference that will become, after the big bang, the ordinary inertial background of the fundamental observers. At finite values of the energy can correspond, however, a finite number of excitations $\nu \neq 0$, the "excitations of the inertial field $\boldsymbol{a}$". In this sense, every point in space is a potential oscillator. The maximum excitation corresponds to $\nu = 137 \approx 1/\alpha$, because in this case we have $T_0 = t_0$. Since in the Einsteinian limit $t_0 \to \infty$ only the value $\nu = 0$ survives, the pre-vacuum molecules (and the related dark matter clusters that we are going to describe) are an effect of the finiteness of $t_0$. Moreover is $u \approx h/k\alpha t_0$, so the ratio between this excess and the background value $\hbar c/k x_0$ (where $0 \leq x_0 \leq 10^{-13}$ cm) is $\leq 10^{-39}$. The link between $T_0$ and the molecular radius $\rho$ is defined by the following relation:

$$\frac{c}{\rho t_0} = \frac{4\pi^2}{T_0^2} \quad (16)$$

that is:

$$\rho = \frac{v^2 \alpha^2 c t_0}{4\pi^2} \quad (17)$$

The quantization of $T_0$ and relations (14), (17) are the basic assumptions of this model.

## 3. "After" the big bang

For $t \approx \theta_0$ the size of the private space of a fundamental observer is $\approx c\theta_0$ (the initial singularity is removed in this description) and the origins of all the frames of reference with $v \neq 0$, which form a finite set, are concentrated in this space. In a point whose distance from the $i$-th of these origins is $x_i$, a test body undergoes a total acceleration given by the sum of all the accelerations $a(x_i)$ relative to each origin, *each computed without taking into account neither R(t) nor the initial contraction for a scale factor $\approx N_D$*; that is, the "archaic" value of the $x_i$ is used. With the increase of $t$ the various contributions $a(x_i)$ overlap less and less due to the expansion of space and they are therefore less and less averaged. There is therefore a gradual (apparent) growth of the dark matter. When the space radius is $ct_0$ the situation existing in the Archaic Universe is restored. The dark matter thus ceases to appear and from then on is simply diluted (this happens roughly for $t \approx 7 \times 10^9$ y ) [15]. This leads to the appearance of more regular morphologies for matter clusters favored by the molecules.

## 4. Dark Matter

Let's now return to consider the single molecule. The field $\boldsymbol{a}$ cannot be outgoing from the origin, because in this case a body subject only to this field (that is, whose acceleration coincides with $\boldsymbol{a}$) would be rejected by the origin and would not have a defined period of oscillation. This period is defined in the interval $0 \leq x \leq \delta$ where it is constant and equal to $T_1$ (for an harmonic oscillator). The fact that acceleration is directed towards the origin makes the molecule a region where the condensation of matter is favored. Within this region the acceleration increases linearly, according to (14), from a minimum value (which must be zero for reasons of symmetry) to a maximum value $a_{max} = 4\pi^2 \delta/T_1^2$. Let us pose $T_1 = T_0 g$. If we indicate with $\sigma\rho$ the value of $x$ at which the extension of the acceleration profile evaluated for $0 \leq x \leq \delta$ equals $c/t_0$, we have:

$$a_{max} = \frac{4\pi^2 \delta}{T_0^2 g^2} \quad (18)$$

$$\frac{c}{t_0} = \int_0^{\sigma\varrho} \frac{4\pi^2}{T_1^2} dx = \frac{4\pi^2 \sigma\varrho}{T_0^2 g^2} \Rightarrow \frac{T_0^2}{4\pi^2} = \frac{\sigma\varrho t_0}{g^2 c} \quad (19)$$

From (19) and (16) it is then obtained immediately $\sigma = g^2$.

Within the sphere $0 \leq x \leq \rho$ the Laplace equation takes the form:

$$\frac{1}{x^2}\frac{\partial}{\partial x}\left(x^2 \frac{\partial u}{\partial x}\right) = 0 \quad (20)$$

That is:

$$u = \frac{C_0}{x} + C_1 \quad (21)$$

Where $C_0$, $C_1$ are constant. For $0 \leq x \leq \delta$ we have $T = h/ku = T_1$, thus it is $C_0 = 0$, $C_1 = h/kT_1$. The solution in the spherical crown $\delta \leq x \leq \rho$ will be obtained by posing $u = h/kT_1$ for $x = \delta$, $u = 0$ for $x = \rho$. We have:

$$u = \frac{h}{kT_1 \left(\frac{1}{\delta} - \frac{1}{\rho}\right)} \left(\frac{1}{x} - \frac{1}{\rho}\right) \quad (22)$$

Remembering that $T = h/ku$ one has:

$$T = \frac{T_1(\rho - \delta)}{\delta \left(\frac{\rho}{x} - 1\right)} \quad (23)$$

For $x \ll \rho$ the following linear approximation holds:

$$T \sim A(x - \delta) + B; \quad A = \frac{T_1(\rho - \delta)}{\delta \rho}; \quad B = A\delta \quad (24)$$

For $0 \leq x \leq \delta$ the equation (18) can be rewritten as:

$$\frac{d}{dx}\left(\frac{a_{max}}{a}\right) = -\frac{\delta}{x^2} \quad (25)$$

Instead, for $\delta \leq x \leq \rho$, we have from (24):

$$a = a_{max} - \frac{4\pi^2}{A}\left(\frac{1}{B} - \frac{1}{A(x - \delta) + B}\right) \quad (26)$$

and thus:

$$\frac{a_{max}}{a} = \frac{1}{1 + \frac{4\pi^2}{Aa_{max}}\left[\frac{1}{A(x - \delta) + B} - \frac{1}{B}\right]} \quad (27)$$

Let us consider, within the limits of linear approximation (24), $x$ values such that $(x - \delta)/\delta \ll ABa_{max}/4\pi^2 = (1 - \delta/\rho)^2$. The further approximation applies to them:

$$\frac{a_{max}}{a} = 1 + \frac{4\pi^2}{a_{max}B^2}(x - \delta) \Rightarrow \frac{d}{dx}\left(\frac{a_{max}}{a}\right) = \frac{4\pi^2}{a_{max}B^2} = \frac{\rho^2}{\delta(\rho - \delta)^2} \quad (28)$$

By posing $\delta = f\rho$, with $0 \leq f \leq 1$, this expression becomes $\lambda/\rho$, where $\lambda = 1/[f(1-f)^2]$; it coincides with $1/\delta$ in the limit $f \ll 1$.

The two expressions obtained respectively for $0 \leq x \leq \delta$ and $\delta \leq x \leq \rho$ (exact the first one, approximate the second one) can be combined in a single expression as follows:

$$\frac{d}{dx}\left(\frac{a_{max}}{a}\right) = \frac{\lambda}{\rho}\Theta(x-\delta) - \frac{\delta}{x^2}[1 - \Theta(x-\delta)] \quad (29)$$

The application of linear approximation to observational data seems to indicate a preference (whose significance is still an open problem) for three cases: (I) completely harmonic case with $\delta \approx \rho$, $\sigma \approx 1$, for galaxy clusters; (II) $\sigma = \delta/\rho = f \ll 1$ for single galaxies; (III) $\sigma = 1$, $f$ not negligible, for single galaxies. These three cases can be summarized, respectively, in the three groups of equations:

$$\begin{cases} a(x) = \dfrac{c}{t_0}\dfrac{x}{\rho} & 0 < x < \rho \\ a(x) = 0 & x > \rho \end{cases} \quad (30)$$

$$\begin{cases} a(x) = \dfrac{c}{t_0}\dfrac{x}{\delta} & 0 \leq x \leq \delta \\ a(x) = \dfrac{c}{t_0}\dfrac{x}{\delta} & \delta \leq x \leq \rho \\ a(x) = 0 & x > \rho \end{cases} \quad (31)$$

$$\begin{cases} a(x) = \dfrac{c}{t_0}\dfrac{x}{\delta} & 0 \leq x \leq \delta \\ a(x) = \dfrac{c}{t_0}\dfrac{\delta}{\rho}\dfrac{1}{1+\lambda\dfrac{x-\delta}{\rho}} & \delta \leq x \leq \rho \\ a(x) = 0 & x > \rho \end{cases} \quad (32)$$

The application of (30), (31), (32) to concrete cases is examined in other works [12,13]; the fitting of the experimental data is overall satisfactory. We note that is $f \approx \sigma$ (or $a_{max} \approx c/t_0$), regardless of the specific case. It is necessary to underline that the existence of a "molecule" is completely independent from the possible presence of baryonic matter inside it; in this context, therefore, thickening of dark matter is possible in areas where baryon matter is absent. In particular, systems such as the "Bullet Cluster" do not seem to pose significant interpretative problems [12].

## 5. Scale Law

Regardless of the specific case and the applicability of the linear approximation it is possible to derive a universal law of scale invariance [13]. Let us consider again the expression of the acceleration for $0 \leq x \leq \delta$ and introduce the dark mass $M(x)$ to the following way:

$$a(x) = \frac{a_{max} x}{\delta} = \frac{c\delta}{\sigma\rho t_0}\frac{x}{\delta} = \frac{cx}{\sigma\rho t_o} = G\frac{M(x)}{x^2} \quad (33)$$

where $G$ is the Newton's gravitational constant. By defining the density $\mu$ of dark matter through the relationship:

$$M(x) = \frac{4}{3}\pi\mu x^3 \qquad (34)$$

and posing $c/t_0 = 6.2\times10^{-10}$ ms$^{-2}$, the following result is immediately derived:

$$\mu(\sigma\rho) = \frac{3c}{4\pi G t_0} = 2.12\frac{Kg}{m^2} = 9\cdot 10^2 \frac{M_{Sun}}{(pc)^2} \qquad (35)$$

where $M_{Sun}$ is the solar mass. That is:

$$log_{10}\left(\frac{\mu\sigma\rho}{M_{Sun}/(pc)^2}\right) = 2.95 \sim 3 \qquad (36)$$

To compare this result with the observations one can take the central density of dark matter of a galaxy as an estimator of $\mu$ and the length $r_0$ of the profile of Burkert as an estimator of $\sigma\rho$ [13]. With these identifications, the obtained relationship is confirmed by extensive observational research related to galaxies of all sizes, from elliptical dwarfs to giant spirals [16]. The validity of this scale law allows to define the parameter $\sigma$ in terms of $\mu$, which is probably a condition expressing an accidental fact, and $\rho$, fixed by the quantum number $\nu$. We conclude by returning to the topic of the relationship between cosmos and particles, which inspired the definition of $T_0$. Let us define the "electron density" $\mu_e$ through the relation:

$$m_e = \frac{4}{3}\pi\mu_e\left(\frac{e^2}{m_e c^2}\right)^3 \qquad (37)$$

where $m_e$ is the mass of electron at rest. As can be seen, a radius equal to the classical radius is attributed to the electron. Then we have, within one order of magnitude, the following numerical coincidence:

$$\mu(\sigma\rho) \sim \mu_e\left(\frac{e^2}{m_e c^2}\right) \qquad (38)$$

In other words, the electron satisfactorily matches the observed scale invariance for galaxies. The last relation can also be deduced theoretically assuming that the ratio between the classical radius of the electron and $ct_0$ is equal to that between its gravitational and electrostatic self-energies. This ratio, as is known from Dirac and Eddington cosmologies, is $\approx 1/N_D$.

## 6. In Search of Quantum Traces

After Rosen paper, a group of Italian scientists has tried to apply his simple scheme of cosmological quantization to explain the apparent regularity in the galaxy distribution [17]. Indeed it seems that the distribution of galaxies is non–random and some unknown mechanism has led to the formation of regularity in the galaxy distribution with a characteristic scale of $128h^{-1}Mpc$ with $0.5 \leq h \leq 1$ [18]. In [17], unlike Rosen, the authors interpret $m$ as the mass of a galaxy and for this reason, in their interpretation, we have that a galaxy of mass $m$ has the probability $|\psi|^2$ to be at a given scale factor $a(t)$. The relation between the scale factor and the red-shift is

$$\frac{\dot{a}}{a} = H = -\frac{\dot{z}}{z+1} \quad (39)$$

so

$$\frac{a_0}{a} = 1 + z \quad (40)$$

we than can obtain the probability amplitude to find a given galaxy at a given red–shift $z$, at time $t$. To complete this analysis, by using the standard quantum mechanics, the Schrödinger stationary equation

$$\mathbb{H}\psi = E\psi \quad (41)$$

presents the stationary state of energy E

$$\psi(a,t) = \psi(a)e^{-\frac{iEt}{\hbar}} \quad (42)$$

Therefore, we can write

$$-\frac{\hbar^2}{2m}\frac{d^2\psi}{da^2} + V(a)\psi = E\psi \quad (43)$$

and, by using equation (7), the Schrödinger stationary equation can be written in the form

$$\psi'' + \left[Aa^{(2-3\gamma)} + Ba^2 + C\right]\psi = 0 \quad (44)$$

where, the prime indicates the derivative with respect to $a$ and

$$A = \frac{8\pi m^2 G}{3\hbar^2 c^2}\epsilon_0 a^{3\gamma}, B = \frac{m^2 \Lambda c^2}{3\hbar^2}, C = -\frac{m^2 k c^2}{\hbar^2} \quad (45)$$

The authors examine the equation (44) in detail and, for mathematical insights, we recommend reading the paper [17]. Indeed the authors show that there are several possibilities to get oscillatory solutions. They depend on the type of cosmic fluid and the type of spatial metrics. For example, in a dust model with cosmological constant, they find solutions in terms of Bessel functions that have an asymptotic behavior in good agreement with the experimental data. Obviously a good result is to get 8 oscillations with periodicity of $128h^{-1}Mpc$ in $2000h^{-1}Mpc$ and that is in a red–shift range from z = 0 to z = 0.5. In our opinion, since the quantum imprinting there has been during the radiaton era with $\gamma = 4/3$ and, furthermore, it seems that space is flat [19], above all it would be interesting to analyze the solutions of

$$\psi'' + \left[\frac{A}{a^2} + Ba^2\right]\psi = 0 \quad (46)$$

More in general, if one considers equation (46) as

$$\psi'' + P\psi = 0 \quad (47)$$

and with the replacement

$$\psi = u \exp\left\{i \int \frac{k_1}{u^2} dx\right\}$$

$$\psi' = \left(u' + i\frac{k_1}{u}\right) \exp\left\{i \int \frac{k_1}{u^2} dx\right\}$$

$$\psi'' = \left(u'' - \frac{k_1^2}{u^3}\right) \exp\left\{i \int \frac{k_1}{u^2} dx\right\}$$

one obtains

$$u'' - \frac{k_1}{u^2} + P u = 0 \quad (48)$$

that in our case it yelds to

$$u'' - \frac{k_1}{u^2} + [Aa^{(2-3\gamma)} + Ba^2 + C]u = 0; \quad u_0 = |\psi_0|; \quad \psi_0 \neq 0; \quad \psi_0' = \frac{\psi_0}{|\psi_0|}\left(u_0' + i\frac{k_1}{|\psi_0|}\right)$$

from Majorana notes [20], this gives a Bessel-like solution

$$\psi = u\left[A_1 \exp\left\{i \int \frac{k_1}{u^2} dx\right\} + B_1 \exp\left\{-i \int \frac{k_1}{u^2} dx\right\}\right]$$

by setting $u_1 = \frac{u}{\sqrt{k_1}}$ one obtains then a similar solution for the following differential equation,

$$u_1'' - \frac{1}{u^3} + u_1 P = 0 \quad (49)$$

$$\psi = u_1\left[A_1 \exp\left\{i \int \frac{1}{u_1^2} dx\right\} + B_1 \exp\left\{-i \int \frac{1}{u_1^2} dx\right\}\right]$$

as the parameter P depends on a, we can analyze the regimes under small/large oscillations. If P is slowly varying, then is obtained an oscillatory solution. This behavior suggests that when the accelerated expansion ends, an oscillatory regime can take place.

Consider as an example the solution $u = P^{-1/4}$ for $P > 0$. An oscillatory solution is immediately obtained:

$$\psi = \frac{1}{\sqrt[4]{P}}\left(A_1 \cos \int \sqrt{P} dx + B_1 \sin \int \sqrt{P} dx\right)$$

instead, this solution explodes when $P < 0$;

$$\psi = \frac{1}{\sqrt[4]{P}}\left(A_1 \exp \int \sqrt{|P|} dx + B_1 \exp \int -\sqrt{|P|} dx\right)$$

when P is slowly varying, then the following condition is obtained:

$$k > \frac{8\pi G}{c^2} \frac{\epsilon_0}{c^2} \frac{a^2}{3} + \Lambda \frac{a^2}{3}$$

This condition implies that:

for $k = 0$ we do not have solutions obeying this condition unless $\Lambda < 0$

for $k = 1$ we can have accelerated expansion if the Universe is closed.

For $k = -1$ the condition is valid only if $\Lambda < 0$

From these results one obtains oscillatory solutions for $P > 0$ and $P$ is slowly varying, a sort of breating mode and solutions with negative cosmological constants[1].

## 7. The Dark Side of the Universe

Over the years, the Universe has become very "dark". Indeed, it is well known that the best description of cosmological evolution is the so called $\Lambda CDM$ model. This model is a consequence of experimental observations of supernova redshifts [21-22], clustering of galaxies [23-25] and Cosmic Microwave Background (CMB) [26-28]. The basic ingredients of $\Lambda CDM$ model are 69.2% of Dark Energy (DE), the cosmological constant that is the cause of the accelerated expansion of the Universe, and 30.8% of self-gravitating matter of which only 4.9% is of ordinary luminous matter while the remaining 25.9% of the energy density would be composed by Dark Matter (DM). The hypothesis of DM is born when C. Zwicky measured the velocity dispersion of the Coma cluster of galaxies [29]. Experimental observations, mainly due to Vera Rubin's research group, clearly show that galactic rotational speeds do not decrease following Keplerian way and the main problem concerns specially areas outside the luminous part of the galaxy. In these areas there are neutral hydrogen clouds at large distances from the rotational centre and they too are moving at constant tangential velocity although, in those regions, the sky is totally dark. A minority of scientists seek to change in various ways the gravitational law or the Newton second law and certainly the most famous attempt is due to the Israeli scientist Mordehai Milgrom. He, from 1983 to now, developed the so called MOND (Modified Newtonian Dynamics) theory in order to explain a variety of astronomical phenomena without requiring the presence of DM [30-31]. Although there are models that try to interpret the experimental results without "dark" components [32-34], almost all of the scientific community believes that DE and DM are inevitabile. For example, in a recent paper [35] the authors, starting from data reported for main sequence stars under the Sloan Digital Sky Survey collaboration, showed that the periodic spectral modulations could be due to DM effects. Returning to the topic of our paper, we must observe that the cosmological Schrödinger equation has also been applied to the case of search for DM candidates and to analyze the origin and nature of DE [36-37]. In [36] the author analyzes the numerical solutions of (44) in two remarkable cases with a different interpretation of the mass. He predicts, in this way, that DM is composed of a quantum particle of very low mass and is clustered around the luminous matter of galaxies. Let us remember the following relations useful for implementing the equation in a mathematical software to find numerical solutions

---

[1] If we also analyse the equation in the form $y'' + \left[\frac{A}{a^2} + Ba^2\right]y = 0$, $P > 0$ is always valid, instead $P < 0$ is never satisfied. In this case we have only oscillatory regimes like in the DM-axion stellar oscillations.

$$\begin{cases} \Omega_M + \Omega_\Lambda + \Omega_k = \frac{\rho_0}{\rho_C} + \frac{\Lambda c^2}{3H_0^2} - \frac{kc^2}{a_0 H_0^2} = 1 \\ \quad H_0 = 100h \frac{Km}{Mpc} \\ a_0 = \frac{c}{H_0} = 3000h^{-1} Mpc \ if \ k = 0 \\ \quad a_0 = \frac{3000h^{-1} Mpc}{\sqrt{\Omega_M + \Omega_\Lambda - 1}} \ if \ k = 1 \\ \quad a_0 = \frac{3000h^{-1} Mpc}{\sqrt{1 - \Omega_M + \Omega_\Lambda}} \ if \ k = -1 \\ \quad \rho_C = \frac{3H_0^2}{8\pi G} \\ \quad \hbar = 1.054 \cdot \frac{10^{-34} J \cdot s}{rad} \\ \quad c = 3 \cdot 10^8 m/s \end{cases} \quad (50)$$

In a recent paper [37], instead, the authors follow Rosen's original paper and that is, $m$, in the equation, is interpreted as the mass of the whole Universe. Moreover, they look at the cosmological constant not as a perturbation of potential energy, but as related to the total energy. Finally, they make a suitable change of variables

$$R = R_0 ln\left(\frac{a}{a_m}\right) \quad (51)$$

where $R_0$ and $a_m < a_0$ are arbitrary constants with

$$a = a_m e^{\frac{R}{R_0}} \quad (52)$$

By posing

$$\begin{cases} T = \frac{1}{2} m \dot{R}^2 \\ V_0 = \frac{4\pi m G a_0^3 \rho_0 R_0^2}{3 a_m^3} \\ E = \frac{m \Lambda c^2 R_0^2}{6} \end{cases} \quad (53)$$

these quantities reduce the cosmological Schrödinger equation in a Bessel-like form

$$\mu^2 \frac{d^2\psi}{d\mu^2} + \mu \frac{d\psi}{d\mu} + \left(\mu^2 + \frac{4K^2}{\alpha^2}\right)\psi = 0 \quad (54)$$

with

$$\begin{cases} \beta^2 = \frac{2mV_0}{\hbar^2} \\ K^2 = \frac{2mE}{\hbar^2} \\ \alpha = \frac{3}{R_0} \end{cases} \quad (55)$$

By fixing a suitable boundary condition on the eigenfunctions of the Schrödinger equation, they calculate the corresponding eigenvalues of energy obtaining that the wave function of the Universe is a Bessel function of purely imaginary order. This result seems very interesting because Bessel functions occur in many branches of mathematical physics and, in particolar, the Bessel functions with purely imaginary order have many applications in physics as you can read in [38]. The most important result is that DE can be seen as the energy of the quantum state of the Universe. The Universe is not in a fundamental state with zero energy, but it is in an excited state with a non-vanishing value of energy.

## Conclusions

In this paper we have reviewed the recent developments of Rosen's approach to quantum cosmology and the concept of Archaic Universe. The first approach is conceptually very simple. The cosmological quantum equation is nothing else but the quantum counterpart of the classical Einstein first order equation. The origin of the cosmological constant is still unknown and, following Rosen quantization, it is possible to explain dark energy, through a complex mathematical formalism, as a simple excited state of the Universe. Furthermore, if we change the interpretation of the mass in the equation, it is possible to use this approach to search for Dark Matter particles. Finally, the oscillating solutions can be connected in natural way to the correlation function, which gives the features of the distribution of superclusters of galaxies. The Archaic Universe, instead, proposes an interpretation of DM and of the morphogenesis linked to a reading of the "big-bang" as nucleation in a vacuum geometrically constrained and observed under particular conditions of projectivity [39]. Nucleation is a quantum process, and DM is a global systemic effect of flocculation of matter. The Archaic Universe is compatible with it but does not necessarily require a "particle candidate". Why should we deal with these theories? In these years the Standard Model has had an increasing number of confirmations and the open questions, at the moment, do not seem to require deep structural changes. Apart from the singular case of Peccei-Quinn-Wilczek-t'Hooft's axion, the reading of DM in particle terms does not seem to be the only one. We believe that theories that explore the possibility of explaining DM in systemic terms, but not based on ad hoc assumptions, albeit brilliant, deserve attention; for example, MOND theories. In particular, the theories here considered, are born as promising models of quantum cosmology that seem to merge gravity and quantum physics in an organic way. Naturally the future of these theories will be decided by a close comparison with observational data and particle models of DM [40-41].